\documentclass[prl,aps,twocolumn]{revtex4}

\usepackage{graphics}
\begin{document}

\title{Fundamentals for immediate implementation of a quantum secured Internet}

\author{Geraldo A. Barbosa$^*$}
%\affiliation{Northwestern University, Department of Electrical
%Engineering and Computer Science, 2145 N. Sheridan Road, Evanston,
%IL 60208-3118 }
\affiliation{QuantaSec--Research in Quantum
Cryptography Ltd., Av. Portugal 1558, Belo Horizonte MG 31550-000
Brazil. }

\date{16 August 06. A typo in the eigenvalues in version V1 was corrected and a few short comments added.}
\begin{abstract}

This work shows how a secure Internet for users A and B can be
implemented through a fast key distribution system that uses
physical noise to encrypt information transmitted in deterministic
form. Starting from a shared secret random sequence between them,
long sequences of fresh random bits can be shared in a secure way
and not involving a third party.  The shared decrypted random bits
-encrypted by noise at the source- are subsequently utilized for
one-time-pad data encryption. The physical generated protection is
not susceptible to advances in computation or mathematics. In
particular, it does not depend on the difficulty of factoring
numbers in primes. Also, there is no use of Linear Feed Back Shift
Registers. The attacker has free access to the communication
channels and may acquire arbitrary number of copies of the
transmitted signal without lowering the security level. No intrusion
detection method is needed.
%{\bf KeyWords:} {\em Cryptography, Physical
%Noise, Internet, Secret Key, No Third-Party, fast communication,
%amplification allowed}

\end{abstract}

\maketitle

\newcommand{\be}{\begin{equation}}
\newcommand{\ee}{\end{equation}}
\newcommand{\bea}{\begin{eqnarray}}
\newcommand{\eea}{\end{eqnarray}}

\section{Introduction}

Are there conditions to have a physical-noise
%(from quantum fluctuations)
secured Internet (NSI)-instead of a relying on the difficulty of
factoring numbers in primes- implemented for common users? The
answer is Yes; and the associated cost is very low -- this paper
shows how it can be immediately implemented.

Let us start with the following classical scenario: User A (Alice)
uses a (private) laptop with no network or web access to encrypt a
message $X$ onto a file $C$ to be sent to end user B (Bob). The
encryption is done with a one-time-pad key sequence of truly random
bits stored in his computer. This random sequence was beforehand
shared with B. File C is fully characterized and transferred to a
computer with web access. From this point on file C is  appended
with all identification information, split into different packets
--as demanded by the Internet protocol (IP) stack \cite{IP}-- and
routed through any available physical link to the end user B . All
modifications are done according to the requirements of the Open
Systems Interconnection (OSI) layers and protocols \cite{OSI}
including particularities demanded by web service providers. Among
these procedures error correction protocols are included. At the
destination, file C is assembled bit-by-bit and delivered to B.
Stripped from communication information attached to it, B checks for
file integrity and decrypt C using the shared random sequence. The
obtained message has unconditionally proven confidentiality and
authentication. These procedures are independent of any established
 Internet security protocol. It is expected that Internet should not
modify the bit content of the original file after its full recover
by B. In case A and B are performing a commercial operation,
non-repudiation is easily assured to both users. The one-time-pad
protocol is not intended to modify the Internet protocols in anyway.
It is a complementary method for secure communication between two
users that works independently and on top of all IP protocols. A and
B have their secure communication totally under their control. This
way, the users know that their protection does not rely on
mathematical difficulties to factoring number in primes nor they
depend on third parties to assure the desired security.

The reader may be saying: But this is a classical encryption between
two users and well known to be secure; there is nothing physical or
new here. However, is this scenario realistic if A and B wishes to
use it continuously? The answer to this question has been negative:
The practical difficulties for A and B to keep sharing long
sequences of one-time-pad keys makes this scenario highly
inefficient and, therefore, useless for most uses. Practical systems
have not yet been devised that allow A and B to use the starting
secret $K_0$ as a support to transfer or exchange fresh sequences of
random numbers.
%One may also
%argue that this difficulty grows exponentially for simultaneous
%communication between many users.

    This work shows a practical way that allows A and B to share through the Internet
    these new sequences of random keys in a secure way. This could bring life to the starting scenario.
    The basic ingredients needed are a very simple software, a physical random generator
    (not a pseudo random generator)
    coupled to the private computer and a starting secret key $K_0$ shared by A and B.
    This starting sequence $K_0$ will seed a long sequence of truly random bits.
    In a sense, it can be seen as a one-time-pad booster: Starting from a shared seed $K_0$,
    A and B end up sharing a sequence of $L \:(>>K_0)$ random bits. No third party is used to
    establish the key sequences to be used by A and B.  No need for intrusion detection
    exists. Furthermore, all of these elements allow immediate
    implementation -- even commercial (truly) random generators exist \cite{IDQuantique} for moderate speeds.

One may also argue that some quantum protocols exist that are proven
secure as the well known single photon protocol BB84 \cite{BB84} to
distribute random bit sequences and, therefore, what novelty is
being offered?
%why to use generated quantum files sent over a deterministic carrier?
Single-photon protocols cannot be amplified and therefore do not
work for the long-haul communications necessary for the Internet.
Furthermore, signals from single-photon protocols cannot be
converted from optical to electrical and back to optical without
loss of security. Nor they are practical for wavelength multiplexing
(WDM). These steps are necessary for the Internet. Alternative
systems such as those using discrete or continuous variable
processes  relying on homodyne measurements (e.g., \cite{grangier})
are very sensitive to noise.  This leads to low key rate transfer
and, even more serious, they cannot work in the naturally disturbed
and complex Internet networks. In this sense, direct quantum
communication over the Internet is not realistic.

    On the other hand, some advantages of this NSI are:
1) The protocol is established at the user level where the secure
data/message is prepared. This message/data is deterministic  and
noiseless but carries, as it will be shown, truly random information
that was generated by a physical process. The key distribution
procedure uses the installed communication network but its
protection depends on physical laws instead of mathematical
complexities. 2) The data is prepared in any convenient form for the
underlying OSI layers with no need for modification of the IP in
use. A simple binary file can be prepared by A or B to be sent
through the OSI stack. These characteristics will not be dependent
of security procedures established at the OSI's ``Presentation''
layer. Normal data manipulation demanded by OSI protocols can be
applied. The only and usually expected requirement is that the end
user receives the data file as it was delivered by the sender,
bit-by-bit. This presupposes use of error correction protocols to
guarantee perfect delivery of the ciphered message to the end user.
The IP protocols in use are then left untouched; just a private
protection layer is added by the users. This added layer, under
user's control, presents no risk in the eventual creation of
algorithms for fast factoring of numbers in primes. Also, even
creation of a quantum computer or quantum processors does not
decrease the physical protection tied to the signals. There resides
the special value of this system and its proposition as a secure
layer for users that demand protection based on physical principles.

The central problem is how to distribute over the Internet secure
sequences of random bits $R_i$. This is the main puzzle with a
solution presented in this work. Before discussing this fundamental
problem and the proposed solution, one may state that if this is
true, it is clear that ciphered messages based on one-time-pad could
be sent over any physical channel with no need to further obscure
the transmission for protection. The cipher message could even be
made public because the protection is guaranteed by the one-time-pad
method itself. As another consequence, no intrusion detection
mechanism will be needed.

Use of classical carriers to carry recorded quantum information is a
normal process that is often not perceived. Scientific journals use
this process constantly --although they do not require a special
protection. Understanding quantum phenomena as sets of probable
events or different possible quantum trajectories, the classical
information obtained from instrumental clicks is nothing more than
recording one amongst the many possible quantum trajectories.
Repetition of the same measurement operation may lead to a very
distinct result; that is to say, a record of another trajectory
among the possible ones. The files to be sent over the Internet are
to be obtained in samplings of single events (bits). The information
protection desired relies on the multitude of possible quantum
trajectories that generates each single bit of the random sequence.
This is completely different from using pseudo noise generated in a
deterministic process (hardware stream ciphers), whose generation
mechanism can be searched, discovered and used by the attacker
\cite{homodyneattack}. One may easily argue, for example, that phase
fluctations on a laser output or in thermal radiation are not
quantum.  Both can be represented by Gaussian random processes.
Several definitions (e.g. Glauber's Positive $P$ representation
\cite{glauber} or Mandel's $Q$ parameter \cite{MandelWolf}) can be
utilized to classify these fluctuations as Poissonian or
super-Poissonian. Light in a coherent state will be at the boundary
between the ``classical'' and ``quantum'' realms.
 Although the question if electromagnetic radiation can be classified
as classical or quantum is probably a philosophical question, the
importance of the uncontrollable or unpredictable physical
fluctuation in both representations is the important aspect to be
utilized in this work. Sometimes the expression ``quantum
fluctuations'' will be utilized by the author to express
fluctuations associated with the light field. The reader may ignore
the ``quantum'' adjective with no harm for ideas presented. A
quantum calculation is often quite adequate to deal with light
fluctuations and will be utilized.

This work will discuss the physically built-in properties of these
secure files. It will also explains why one-time-pad keys can be
created and shared by A and B through the Internet. This is
basically about physics, not a discussion about software or
deterministic (pseudo-random) stream cipher hardware. The security
is intrinsically connected with fluctuations of the light field.
Before discussing the information content in this process, the
Section \ref{The distribution protocol} will describe the basic
standard distribution protocol. After this description, it is
explained how the recorded files carry the noise protection.

\section{The distribution protocol}
\label{The distribution protocol}

The deterministic signals going through arbitrary communication
channels are encrypted by random signals obtained from optical
sources and are described by non-orthogonal $M$-ry bases.
Distribution of secure data \cite{alphaeta1,alphaetaEXP} and key
distribution over optical channels using $M$-ry bases has been
discussed on recent publications \cite{mykey,infoth,yuenkanter}. The
security of the key distribution process described here relies on a
few points: 1) A shared secrecy by users A and B on a starting key
sequence  $K_0$ and 2) a bit-by-bit uncertainty  Nature-made noise
associated to each bit and recorded on a interleaved  $M$-ry
non-orthogonal bases.

In short, knowledge of  $K_0$ gives the legitimate users the first
mapping of the bases generated by the emitter and allow B to recover
each bit inscribed on every basis used. Sequences of fresh random
bits, by its turn, will be generated by a truly random process and
sent one-by-one between users A and B. Subsequent privacy
amplification procedures statistically exclude the eventually
compromised fraction of shared bits. The batch of secure shared
secret bits (distilled bits) will be used for one-time-pad
encryption. The physical noise from the bit generator protects each
bit from the attacker E (Eve) and provides the information security
level associated with all $R_i$.

 The signals associated to the key sequences $R_i$ are created by
 a physical random generator(PhRG). The noise $N_i$ associated with
 each bit $R_i$ inscribed onto the $M$-ry nonorthogonal basis ($M\geq 2$)
 produces the uncertainty seen by the attacker. This implies that the emitter
has to be equipped to detect and record the signals generated by the
PhRG. In other words, the definition of the measuring system is made
by the emitter, not the attacker. The signal sent is the signal
controlled and measured by the emitter with a detection system of
his choice. No restrictions are placed on the attacker to obtain the
exchanged signals on a public channel. Perfect copies of the
transmission signals can be made public.
%The signals used by the legitimate user obey constraints imposed to provide full security.
Among the properties of the proposed system are: 1) Any public
channel may be used for transmission (optical fibers, TV, microwave,
and so on); 2) The deterministic signals can be amplified with no
security loss; 3) Signals can be converted from electromagnetic to
electrical and back to electromagnetic with no security loss; 4)
Wavelength multiplexing is allowed on the network; 5) Current
Network and IP protocols can be used with no modifications for users
in any IP classes.

\begin{figure}
\centerline{\scalebox{0.47}{\includegraphics{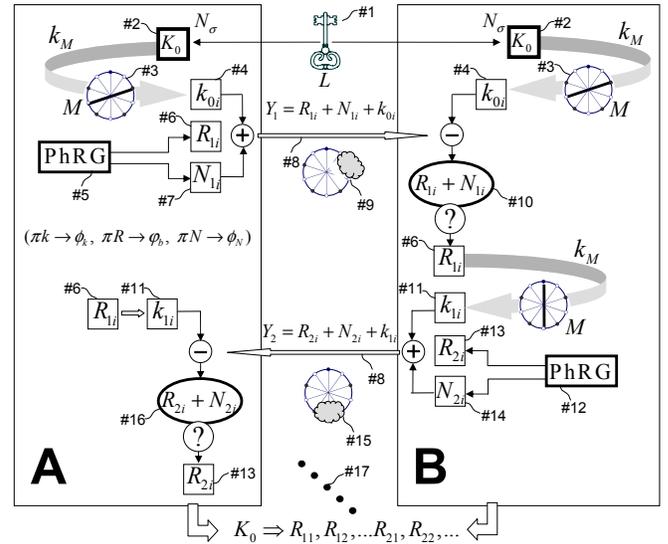}}}
\caption{A sketch of one cycle of operations of the key distribution
process in the Noise Secured Internet is shown. }
\label{InternetBlock}
\end{figure}
Fig. \ref{InternetBlock} shows a block diagram for one cycle of the
key distribution system.
%It describes how legitimate users A and B
%distribute or transfer fresh random key bits generated by a PhRG (to
%be discussed ahead).
Just to describe the protocol and make contact with some of the
available literature (\cite{mykey} to \cite{yuenkanter}) on $M-$ry
cryptography, a description starting with a $M-$ry system of levels
uniformly distributed on the phase circle will be presented. At the
end,
 the $M-$ry system will be simplified to $M=2$ for a speed-up in the
 communication process with no security loss.

 \subsection{The protocol}

 A and B share a starting random key sequence (\#1) designated by $K_0$ (\#2) of
length $L$ (See Fig. \ref{InternetBlock}). These $L$ bits  are
divided into blocks of size $k_M$ ($b(k_M), b(k_{M-1}), ... b(k_1)$)
and each block defines randomly a basis $k_{0i}$ over a
nonorthogonal set of bases. As an example, a uniformly distributed
set of bases can be used, being described on a ciphering wheel (\#3)
\cite{mykey} with $M$ bases, where $M=2^{k_M}$.
\begin{eqnarray}
\label{basis}
 k_{0i}=b(k_M) 2^{k_{M-1}}+b(k_{M-1})
2^{k_{M-2}}+...b(k_1) 2^0\:\:.
\end{eqnarray}
The phase values defining each basis are then given by
\begin{eqnarray}
\label{wheel}
 \phi_{k_{0i}}=\pi \left[
\frac{k_{0i}}{M}+\frac{1-(-1)^{k_{0i}}}{2} \right],
\:k_{0i}=0,1,...M-1.
\end{eqnarray}
In these bases, a bit 1 will be inscribed displaced by $\pi$ with
respect to bit 0 over each basis.

A PhRG (\#5) generates random bits $R_{1i}$ (\#6) that A would like
to transfer securely to B. These signals contain noise $N_{1i}$
(\#7) with a natural phase distribution (e.g., noise inherent to
coherent states) of width $\sigma_{\phi}$. $R_{1i}$ can be
understood in phase units (rd): values 0 or $\pi$ for bits 0 and 1.
For mesoscopic coherent states this noise is appproximately Gaussian
distributed with width $\sigma_{\phi}$ (set such that
$\sigma_{\phi}< \pi/2$).
%$\sigma_{\phi}$ may be written
%$\sigma_{\phi}=\pi N_{\sigma_{\phi}}/M$ where $N_{\sigma}$ is the
%number of bases covered by $N_i$ (See Ref. \cite{mykey}).
The signal to be sent over the generic Internet communication
channel (\#8) (network and servers) is $Y_1=R_{1i}+N_{1i}+k_{0i}$.
The combined effects of $N_{1i}+k_{0i}$ is to hide the bit value
$R_{1i}$ on the ciphering wheel (\#9). Although containing random
information $Y_1$ is a deterministic signal and as such can be
amplified and converted into different signals through arbitrary
Internet nodes without any loss of security.

B has to extract $R_{1i}$ from $Y_1$. To this end he utilizes the
same sequences from $K_0$ utilized by A to generate the base values
$k_{0i}$ (\#4). He subtracts this value from $Y_1$ and obtains
$R_{1i}+N_{1i}$ (\#10) and obtain signals in {\em binary} bases
(single $k_i$ value). The effect of the noise $N_{1i}$ on B's {\em
binary} basis is negligible because $\sigma_{\phi}< \pi/2$ and his
decision on the bit value is easy; therefore, he obtains $R_{1i}$
(\#6). From the received sequence $R_i$ he forms bit blocks of
length $k_M$ and constructs a new base sequence $k_{1i}$. The next
steps are similar to the first ones. Bob's PhRG (\#12) generates
signal containing bits $R_{2i}$ (\#13) associated to noise $N_{2i}$
(\#14). The signal $Y_2=R_{2i}+N_{2i}+k_{1i}$ is sent over the
communication channel (\#8). The bit value $R_{2i}$  is hidden by
the overall noise $N_{2i}+k_{1i}$ (\#15). From her knowledge of
$R_{1i}$ (\#6)  and, therefore, $k_{1i}$ (\#14), Alice subtracts
$k_{1i}$ from $Y_2$ and obtains $R_{2i}+N_{2i}$ (\#16). On her {\em
binary} basis she easily obtains $R_2i$ (\#13). The first cycle is
complete. A and B continue to exchange random sequences as in the
first cycle. The shared sequences ($R_{1i},...,R_{2i},...$), after
privacy amplification, are the random bits to be subsequently
utilized for one-time-pad cipher.

Note that while for noiseless signals $Y_1=b$ and $Y_2=b$ carrying a
repeated bit $b$, one has $Y_1\oplus Y_1=0$, noisy signals give
$Y_1=b+N_1$ and $Y_1=b+N_2$ and, therefore, $Y_1\oplus
Y_1=N_1+N_2(=0\:\:\mbox{or}\:\:1)$. This frustrates correlation
attacks and algebraic attacks constituted of addition-mod2 between
bits. These attacks are efficient against pseudo random encrypted
signals in a noiseless carrier.

\section{The physical random generator}

The random generator is the principal equipment needed to implement
NSI for users A and B. After a brief description of a possible
random generator, its physical aspects will be discussed. For secure
transmission of signals physical  randomness  is necessary because
no known mathematical algorithm has been proven to generate true
random numbers. Several physical sources may be used to this end
such as optical or thermal sources. Optical sources can be much
faster than the thermal ones and are therefore necessary when speed
is required. It is important to say that commercial truly random
generators already exists for moderate speeds \cite{IDQuantique}
what makes immediate NSI implementations possible.
\begin{figure}
\centerline{\scalebox{0.37}{\includegraphics{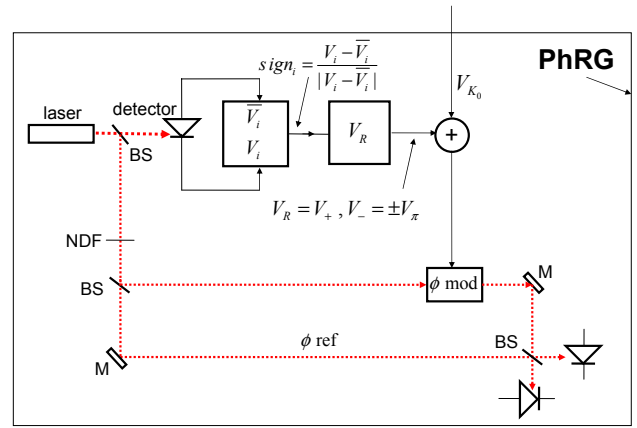}}}
\caption{Sketch of PhRG with a coherent light source. This modulus
can work internally or externally to a computer. The laser beam is
divided by a beam splitter BS. The upper part shows a detecting
system where signals $V_i$ are generated corresponding to the sign
of the generated signal with respect to the average signal
intensity. The laser beam is adjusted to an adequate intensity by a
neutral density filter (or automatized filter). Voltage values
$V_{K_0}$ defining $M$-ry phase bases (e.g, $M=2$) are added to
$V_R$ and applied to the phase modulator. } \label{PhRG}
\end{figure}
Fig. \ref{PhRG} shows a sketch of a PhRG with a coherent light
source modulus. While several design variations are possible, the
PhRG shown can achieve fast speeds compatible with optical channels.
The laser beam is divided by a beam splitter BS. The upper part
shows a detecting system where signals $V_i$  are generated
corresponding to the sign of the generated signal with respect to
the average signal intensity. These binary signals are converted
into binary voltages $V_R=\pm V_{\pi}$ that constitute fresh random
bits to be shared by A and B. The bottom part shows an
interferometer with an optical phase modulator ($\phi$ mod) in one
of the arms. Added voltages $V_{K_0}+V_R$ are applied to the phase
modulator. This way bits are created in randomly chosen
non-orthogonal bases. Detectors at the interferometer output produce
the phase signals carrying basis, bit and noise information shown in
Figure 1 as $Y_i$. These values are automatically recorded and carry
analog information that may be transmitted in binary form.  The
phase uncertainty is approximatelly given (see Refs. \cite{mykey}
and \cite{infoth}) by the Gaussian distribution
\begin{eqnarray}
p_u \simeq e^{- (\Delta \phi)^2 /2 \sigma_{ \phi}^2}\:\:,
\end{eqnarray}
where $\sigma_{\phi}=\sqrt{2/\langle n \rangle}$ and $\langle n
\rangle$ is the average number of photons in one bit. Availability
of PhRG  modules in public places like a cybercafe may be convenient
and less costly for many users. They may generate and record on
portable memories a batch of secure keys or use them to exchange
one-time-pad ciphered information.

\section{Simplified bases}

\begin{figure}
\centerline{\scalebox{0.35}{\includegraphics{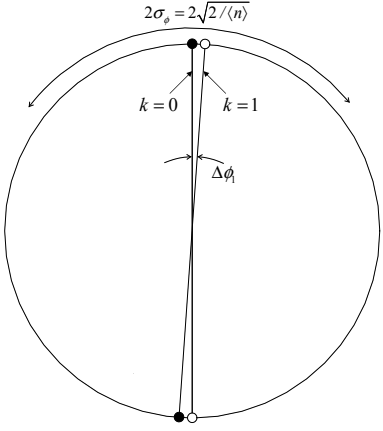}}} \caption{A ciphering set of bases in a phase sector with $M=2$.
$\sigma_{\phi}$ is the standard deviation in the phase caused by fluctuations in the light field. $\Delta \phi_1$ is the
spacing between two bases and should be kept  $\Delta \phi_1\ll \pi/2$.  $\langle n \rangle$ is adjusted so that $\pi/2
> \sigma_{\phi}\gg \Delta \phi_1$. Two states or bits can be inscribed on each basis. Dark
circles indicate positions for a bit 0 and open circles give possible positions for a bit 1.} \label{M2Sector}
\end{figure}
Use of a non-uniform set of bases leads to a more economical system:
instead of the uniformly spaced circle of phases given by Eq.
(\ref{wheel}) one may use just a sector of phase values where the
number of bases is just $M=2$. See Fig. \ref{M2Sector}.
  $\Delta \phi_1$ is the space
between two bases and should be kept  $\Delta \phi_1 \ll \pi/2$.
$\langle n \rangle$ is adjusted so that $\pi/2 >\sigma_{\phi} \gg
\Delta \phi_1$. Two states or bits can be inscribed on each basis.
While only two possible states are to be written on the same basis
(binary states), one should recall that the noise added may require
other angle positions (going to a continuum as necessary) to be
experimentally recorded as well. A simple procedure could be placing
the recorded signal (representing bit, basis and noise) always equal
to the nearest phase position. This way, just the phase values shown
on Fig. \ref{M2Sector} will be needed.

Phase positions in the $s^{\mbox{\tiny th}}$-sequence of random bits
on this $M=2$ sector are given by
\begin{eqnarray}
\label{angle}  &&\phi_{s,i}=(1-R_{s-1,i})\pi\left[ \frac{
1-(-1)^{R_{s,i}}     }{2}
 \right]+\nonumber\\
&&+R_{s-1,i} \left(  \pi \left[ \frac{    1+(-1)^{R_{s,i}}    }{2}
\right]+\Delta \phi_1
 \right), \:k_{0i}=0,1.
\end{eqnarray}
$i$ gives the $i^{\mbox{th}}$ term in the $s$-sequence of length
$L$. The first random sequence $R_{0,i}$ is the shared sequence
$K_0$.

\section{Light signals and information}

Signals to be generated by the PhRG are phase modulated
 coherent signals. A specific $\phi_k$ modulation will be described
by
\begin{eqnarray}
|\alpha e^{i \phi_k}\rangle=e^{-|\alpha|^2/2}\sum_{n=0}^{\infty}
\frac{\left(\alpha e^{i \phi_k}\right)^n}{\sqrt{n!}} |n\rangle\:\:.
\end{eqnarray}
Assume positive $\alpha=|\alpha|$. On the $M=2$ sector shown in Fig.
\ref{M2Sector} the phase modulation angles are
$\phi_k=(0,\Delta\phi_1,\pi,\pi+\Delta\phi_1)$. Each one of these
values is randomly by $R_i$ according to Eq. \ref{angle}.
Statistically, the allowed values are equally distributed with 1/4
probability for each of the angles. It is easy to describe the
desired properties utilizing a quantum formalism. The density matrix
$\widehat{\rho}$ describing these possibilities is
\begin{eqnarray}
\label{rho} \widehat{\rho}&=&\frac{1}{4} \left(  | \alpha e^{i 0}
\rangle \langle \alpha e^{i 0}  | + | \alpha e^{i \Delta\phi_1}
\rangle \langle \alpha e^{i\Delta\phi_1} | \right.\nonumber\\&&
\left.+| \alpha e^{i \pi} \rangle \langle \alpha e^{i\pi} | +|
\alpha e^{i (\pi+\Delta\phi_1)} \rangle \langle \alpha
e^{i(\pi+\Delta\phi_1)} | \right)  \:\:.
\end{eqnarray}
From now on the notation $|\phi_i
\rangle=(|0\rangle,\:|\Delta\phi_1\rangle,\:|\pi\rangle,\:|\pi+\Delta\phi_1\rangle)$
will be used for the modulated coherent states. As the interest is
on small angular separation $\Delta \phi_1$, one may write the
matrix elements of $\widehat{\rho}$ up to the first order
${\cal{O}}(\Delta\phi_1^1)$:
\begin{eqnarray}
\label{rho1}\langle \phi_i|\widehat{\rho}|\phi_k\rangle=\nonumber
%\hspace{7cm}
\\\frac{1}{4} \left(
\begin{array}{cccc}
1&\widehat{\rho}_{0,\Delta \phi_1}&\widehat{\rho}_{0,\pi}& \widehat{\rho}_{0,\pi+\Delta \phi_1}\\
\widehat{\rho}_{\Delta \phi_1,0}&1&\widehat{\rho}_{\Delta \phi_1,\pi}&\widehat{\rho}_{\Delta \phi_1,\pi+\Delta \phi_1}\\
\widehat{\rho}_{\pi,0} &\widehat{\rho}_{\pi,\Delta \phi_1}&1&\widehat{\rho}_{\pi,\pi+\Delta \phi_1}\\
\widehat{\rho}_{\pi+\Delta \phi_1,0}&\widehat{\rho}_{\pi+\Delta
\phi_1,\Delta \phi_1}&\widehat{\rho}_{\pi+\Delta \phi_1,\pi}&1
 \end{array}\right)
\end{eqnarray}
where
\begin{eqnarray}
\widehat{\rho}_{0,\Delta \phi_1}=\widehat{\rho}_{\pi,\pi+\Delta
\phi_1}=\left(1+i |\alpha|^2 \Delta\phi_1 \tanh(
2|\alpha|^2)\right),\nonumber\\\widehat{\rho}_{\Delta
\phi_1,0}=\widehat{\rho}_{\pi+\Delta \phi_1,\pi}=\left(1-i
|\alpha|^2 \Delta\phi_1 \tanh( 2|\alpha|^2)\right),
\end{eqnarray}
and all other terms ($i\neq k$) are equal to
$\widehat{\rho}_{ik}=\mbox{sech}( 2|\alpha|^2)$.
 Diagonalization of $\langle
\phi_i|\widehat{\rho}|\phi_k\rangle$ gives the eigenvalues
$\lambda_i$ and orthonormal eigenstates $|\Psi_i\rangle$ (up to
order $\Delta\phi_1$):
\begin{eqnarray}
\lambda_1&=&0,\\
|\Psi_1\rangle&=&  \frac{1}{\cal N}
\left[  e^{i \phi_C}|0\rangle- |\Delta\phi_1\rangle- e^{i \phi_C}|\pi\rangle+|\pi+\Delta\phi_1\rangle)\right]\nonumber\\
\lambda_2&=&\mbox{sech}( 2|\alpha|^2)\sinh^2( |\alpha|^2),\\
|\Psi_2\rangle&=&\frac{1}{\cal N}
\left[  -e^{i \phi_C}|0\rangle- |\Delta\phi_1\rangle+ e^{i \phi_C}|\pi\rangle+|\pi+\Delta\phi_1\rangle)\right]\nonumber\\
\lambda_3&=&0,\\|\Psi_3\rangle&=&\frac{1}{\cal N}
\left[  -e^{i \phi_T}|0\rangle+ |\Delta\phi_1\rangle- e^{i \phi_T}|\pi\rangle+|\pi+\Delta\phi_1\rangle)\right]\nonumber\\
\lambda_4&=&\frac{1}{2}\left( 1+\mbox{sech}(
2|\alpha|^2)\right),\\|\Psi_4\rangle&=& \frac{1}{\cal N} \left[ e^{i
\phi_T}|0\rangle+ |\Delta\phi_1\rangle+ e^{i
\phi_T}|\pi\rangle+|\pi+\Delta\phi_1\rangle)\right]\nonumber\:\:,
\end{eqnarray}
where
\begin{eqnarray}
&&\lambda_2+\lambda_4=1, \:\:{\cal N}=2\sqrt{2(1-e^{- \langle n
\rangle})},\nonumber\\
&&\phi_C=\arctan\left[ \langle n\rangle \Delta \phi_1 \coth \langle
n\rangle\right]\:\:\mbox{and}\nonumber\\
&&\phi_T=\arctan\left[ \langle n\rangle \Delta \phi_1 \tanh \langle
n\rangle\right]\:\:.
\end{eqnarray}
 The
eigenvalues give the probability of occurrence of the states
$|\Psi_i\rangle$. Due to the non-orthogonality of the bases used, a
modulated state $|\phi_i \rangle$ has projections on all eigenstates
$|\Psi_i\rangle$.

\subsection{Von Neumann and Shannon entropies}

\begin{figure}
\centerline{\scalebox{0.45}{\includegraphics{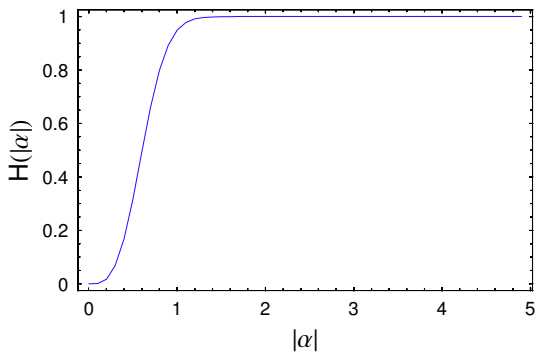}}}
\caption{Von Neumann entropy showing the fast transition from the
quantum regime to the classical bit regime. $|\alpha|=\sqrt{\langle
n \rangle}$, where $\langle n \rangle$ is the average number of
photons per bit signal.} \label{entropy}
\end{figure}

Statistically, the Von Neumann entropy $H(\alpha)$ associated with
the random bits is given by the eigenvalues of $\widehat{\rho}$:
\begin{eqnarray}
H(\alpha)=-\lambda_2 \log_2\lambda_2-\lambda_4 \log_2\lambda_4\:\:.
\end{eqnarray}
%In first order no dependence on $\Delta\phi_1$ exists.
Fig. \ref{entropy} shows $H(\alpha)$ as a function of the coherent
amplitude $|\alpha|$. It is interesting to see that for very small
amplitude $|\alpha|$ (or small number of photons $|\alpha|^2=\langle
n \rangle<1$) the signal carries less than one bit information. Four
states can be used and two of them describes the same bit (two bits
in the same basis). As the probability to sent one of the states is
$1/4$, the
 probability to have one of the two bits sent is $2\times1/4$. Consistently, this
gives the maximum Shannon entropy (as the classical limit of Von
Neumann's entropy) $H_S=2\times (1/2)\log_2 (1/(1/2))=1.$

\subsection{Phase distribution}

The experimental determination of phase in the quantum regime has
been subject of intense study and controversies --for a short
review, see \cite{MandelWolf}. Mesoscopic and classical states are
established with less controversy. Ref. \cite{PeggBarnett}
introduced simple definitions for phase state and phase
distributions that have been frequently used. They will be adopted
here. Thus, in terms of number state bases, the definition of phase
state will be
\begin{eqnarray}
|\phi\rangle=\frac{1}{\sqrt{q+1}}\sum_{n=0}^q e^{i n \phi}
|n\rangle\:\:,
\end{eqnarray}
where $q$ is the number of states taken on a truncated space of the
oscillator Hilbert space \cite{PeggBarnett}. It leads to a classical
phase state for large $q$ and it is quite  adequate for numerical
calculations. Introduce a discrete, orthonormal and complete set of
these states
\begin{eqnarray}
\label{discretephase} \phi_{dm}=\frac{2 \pi m}{q+1}\:\:(m=0,1,...q),
\end{eqnarray}
defined up to an arbitrary fixed reference phase value (The index
$d$ is just to identify the discrete character of this phase). Thus,
$\langle \phi_{dk}|\phi_{dl}\rangle=\delta_{kl}$. A phase operator
is defined by
\begin{eqnarray}
\widehat{\phi}=\sum_{n=0}^q \phi_{dm} |\phi_{dm}\rangle \langle
\phi_{dm}|\:\:.
\end{eqnarray}

\begin{figure}
\centerline{\scalebox{0.55}{\includegraphics{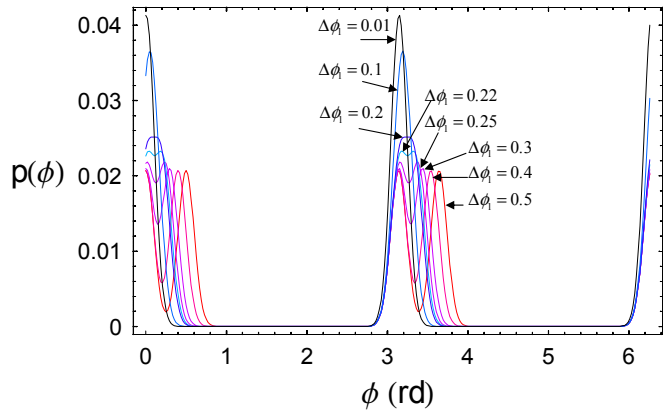}}}
\caption{Phase distribution for $\langle n \rangle=25$ as a function
of $\Delta \phi_1$ values. For large values of $\Delta \phi_1$ the
decision over $\pi$ or $\pi+\Delta \phi_1$ is easily made. For small
 $\Delta \phi_1$ the two distributions merge together. $q=300$ was used to truncate Eq. (\ref{phidistr}).} \label{phiDistribution}
\end{figure}

Given a density operator $\widehat{\rho}$, the phase distribution
$p(\phi)$ is obtained as
\begin{eqnarray}
\label{phasedistribution} p(\phi_{dm})=    \langle \phi_{dm}|
\widehat{\rho}|\phi_{dm}\rangle\:\:,
\end{eqnarray}
with normalization $\sum_{m=0}^q p(\phi_{dm})=1$. From the density
matrix, Eq. (\ref{rho}), the phase distribution
(\ref{phasedistribution}) for all possible realizations of the phase
assignments can be calculated. It gives
\begin{eqnarray}
\label{phidistr} p(\phi_{dm})&=&\frac{e^{-\langle n \rangle}}{4
(q+1)} \sum_{\phi_i}
\left[\left(\sum_{n=0}^q\frac{|\alpha|^m}{\sqrt{n!}}\cos
 n(\phi_i-\phi_{dm})\right)^2\right.
 \nonumber\\
 &+&\left. \left(\sum_{n=0}^q\frac{|\alpha|^m}{\sqrt{n!}}\sin
 n(\phi_i-\phi_{dm})\right)^2\right]\:\:,
\end{eqnarray}
where $\phi_i=(0,\Delta\phi_1,\:\pi,\:\pi+\Delta\phi_1)$ and $m$ is
the phase index introduced in Eq. (\ref{discretephase}). Fig.
\ref{phiDistribution} shows probabilities for occurrences of phases
assigned by the sender. Fig. \ref{phiDistribution} illustrates phase
distributions for a set of $\Delta \phi_1$ and $\langle n
\rangle=25$. Large $\Delta \phi_1$ values imply that recognition for
the attacker of the basis used by the user is easy and leads to bit
recover. On the other hand, for small $\Delta \phi_1$ the linewidth
well exceeds it.

A phase recorded by the sender is sent to the end user and assumed
recorded by the attacker as well. Recovering the bit sent is the aim
of both the end user and the attacker's. To the end user, bit
recovered is easy because it is just a decision between angle ranges
$(-\pi/2,\pi/2)$ or from $(\pi/2,3 \pi/2)$. The attacker, not
knowing the basis used, has to decide between a  phase value or the
neighbor phase, distant from it by $\Delta \phi_1$.

\subsection{Signal-to-noise ratio for phase measurements}

A measure of the attackers ability to recover a phase $\phi_i$ sent
is given by the fundamental signal-to-noise ratio expressed by
\begin{eqnarray}
SNR_{\phi_i}=\frac{\langle  \phi_i|\widehat{\phi} |\phi_i \rangle
^2}{\langle \phi_i|\widehat{\phi}^2 |\phi_i \rangle- \langle\phi_i|
\widehat{\phi} |\phi_i\rangle^2}\:\:.
\end{eqnarray}
The phase expected value $\langle \widehat{\phi} \rangle$ and
$\langle \widehat{\phi}^2 \rangle$ are given by
\begin{eqnarray}
%&&
\langle  \phi_i|\widehat{\phi} |\phi_i \rangle=4
 \sum_{m=0}^q \phi_{dm} p(\phi_{dm})_{\phi_i}\:\:,
 \end{eqnarray}
 and
 \begin{eqnarray}
\langle  \phi_i|\widehat{\phi}^2 |\phi_i \rangle=
 4 \sum_{m=0}^q \phi_{dm}^2 p(\phi_{dm})_{\phi_i}\:\:.
\end{eqnarray}
$ p(\phi_{dm})_{\phi_i}$ is the $\phi_i$ contribution to $
p(\phi_{dm})$. The attacker, E, cannot succeed for $SNR_{\phi_i}\leq
1$. It should be emphasized that the attacker does not have the
capability to perform measurements on the PhRG output. She obtains
single records sent by the user. Not even an ensemble of data for
each bit is sent by the user. A single recording of a single
measurement performed by the user's instruments is the only data
available to the attacker.
\begin{figure}
\centerline{\scalebox{0.55}{\includegraphics{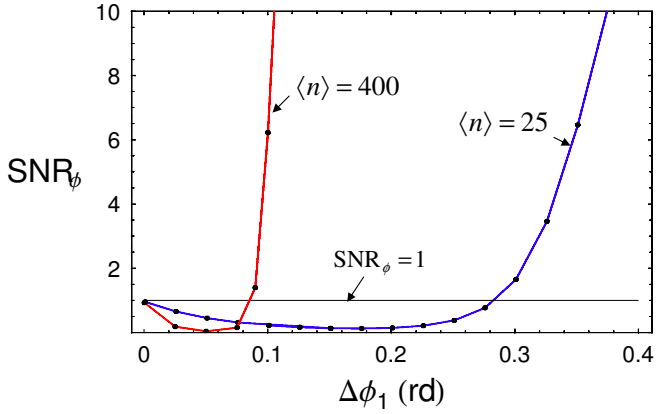}}}
\caption{Signal-to-noise ratio for phase angle as a function of the
separation angle $\Delta \phi_1$ for two values of $\langle
n\rangle$, $25$ and $400$.
%The quantum limits $\Delta
%\phi_{SQL}=1/\sqrt{N}$ and $\Delta \phi_H=1/N$ are shown for each
%case.
} \label{SNRphi}
\end{figure}
Fig. \ref{SNRphi} shows the signal-to-noise ratio $SNR_{\phi_i}$ for
$\langle n \rangle=25$ and $\langle n \rangle=400$ as a function of
$\Delta \phi_1$. It is seen that for a given $\langle n \rangle$ a
small range of $\Delta \phi_1$ values satisfy $SNR_{\phi_i}\leq 1$.
Within this range, the attacker cannot succeed to obtain the correct
bit values (or corresponding phase values). His probability of error
by guessing over the recorded data will be 1/2.

%The conventional quantum limits for phase measurements are also
%shown in Fig. \ref{SNRphi}: the standard quantum limit (SQL),
%$\Delta \phi_{SQL}=1/\sqrt{N}$ and the Heisenberg limit (H), $\Delta
%\phi_H=1/N$.

\section{Attacks}

 One may  wonder about the
cost of a brute force attack to determine the starting key $K_0$
from the transmitted signals. Under the assumption that the
uncertainty presented to the attacker cover some of the bases, the
attacker would know that the basis $k_i$ used in a given
transmission is around a given region within the uncertainty
$N_{\sigma}$.

For the $M$-ry system of uniformly spaced bases this amounts that
only a set of less relevant bits $b_k$ hide the correct basis.
These $b_k$ bits could be permutated in $b_k!$ ways. As each bit
could be either 0 or 1 the total number of permutations to be
searched for each bit emission would be $(\log_2
N_{\sigma})!N_{\sigma}$. For the total number of bits the number of
combinations would be
\begin{eqnarray}
C=2^{K_0}(\log_2 N_{\sigma})! N_{\sigma}\:\:.
\end{eqnarray}
Under this example of a uniform ciphering wheel exemplified by Eq.
(\ref{wheel}), it is understood that the attacker may know the
fraction $1-(N_{\sigma}/M)$ of the total number of shared bits $k_M$
used by A and B to cipher a fresh generated bit. For a sequence of L
shared bits, Eve may obtain $L [1-(N_{\sigma}/M)]$ bits among $L$
because they were not covered by noise. An attack on the key cannot
succeed due to simple reasons: $K_0$ can be chosen with a size that
makes direct search computationally unfeasible (exponential
complexity in $K_0$). After exchange of each random sequence $R_i$
--equally long as $K_0$-- privacy amplification procedures will be
applied, leading to a shorter random sequence for one-time-pad.
 One should stress in this key distribution procedure the starting
key $K_0$ is never to be open to the attacker. This eliminates any
possibility for E to explore correlations between $K_0$ and the
distilled keys after privacy amplification. In fact, $K_0$ can be
destroyed after being used. Therefore, applying key-search trials
for ciphertext decryption on a known-plaintext attack is doomed due
the attacker's computational capability.

For the $M=2$ system all neighbouring levels are covered by noise
($N_{\sigma}\geq 2$). For this case, the same $C\sim2^{K_0}$ makes
unfeasible a brute force attack.

\section{Conclusions}

It has been shown that Internet users will succeed in generating and
sharing, in a fast way, a large number of secret keys to be used in
one-time-pad encryption. They have to start from a shared secret
sequence of random bits   and have a ``hardware'' modulus (Physical
Random Generator-PhRG) added to their computers.
 The
physical noise level is adjusted to hide the random bits being sent.
 Although the
transmitted signals could be openly accessed,  physical noise
inherent to  these signals provide the protection.  No intrusion
detection method is necessary. Privacy amplification protocols
(dependent on the $M$-ry system used) eliminate any fraction of
information that may have eventually obtained by the attackers. As
the security is not based on protocols supported by mathematical
complexities in current use, the security is not dependent on the
difficulties of factoring large numbers in their primes. It was then
shown that by sharing  secure secret key sequences and subsequent
data encryption a secure Internet can be practically implemented.
The system can be easily adjusted to follow any computational
advance while providing security. The random generator works at
optical speeds and the system does not require special Internet
communication protocols. Any network in current use is adequate for
this kind of operation. This system is proposed as a
possible new paradigm for a secure Internet.\\\\
$^*$E-mail: GeraldoABarbosa@hotmail.com

%E. Corndorf, P. Kumar, C. Liang, G. A.  Barbosa, H. P. Yuen,  in R.
%E. Meyers, Y. Shih (Eds.), Proceedings of the 2003 SPIE Annual
%Conference, San Diego, CA, Vol. 5161, 2004, pp. 310 to 319.

\end{document}